\pdfoutput=1
\documentclass[acmsmall,nonacm]{acmart}
\renewcommand\footnotetextcopyrightpermission[1]{}
\setcopyright{none}
\acmDOI{}
\acmISBN{}
\usepackage{booktabs}
\usepackage{tabularx}
\usepackage{multirow}
\usepackage{enumitem}
\usepackage{microtype}
\usepackage{xspace}
\usepackage{xurl}
\usepackage{listings}
\usepackage{subcaption}
\newcolumntype{Y}{>{\raggedright\arraybackslash}X}
\newcolumntype{L}[1]{>{\raggedright\arraybackslash}p{#1}}
\newcommand{\vulvalidate}{\textsc{VulValidate}\xspace}
\newcommand{\ourrq}[1]{\textbf{RQ#1}}
\newcommand{\supp}[1]{Supplementary Section~S#1}
\newcommand{\tablebody}{\normalfont\normalsize\setlength{\tabcolsep}{4pt}\renewcommand{\arraystretch}{1.1}}

\newcommand{\parag}[1]{\smallskip\noindent\textbf{#1}\ \ }
\begin{document}
\title[VulValidate: Auditing Vulnerability Labels]
{\vulvalidate: Auditing Function-Level Vulnerability Labels with Executable Evidence}
\author{Leizhen Zhang}
\affiliation{%
  \institution{University of Louisiana at Lafayette}
  \city{Lafayette}
  \state{Louisiana}
  \country{USA}
}
\email{leizhen.zhang1@louisiana.edu}

\author{Sheng Chen}
\authornote{Corresponding author.}
\affiliation{%
  \institution{University of Louisiana at Lafayette}
  \city{Lafayette}
  \state{Louisiana}
  \country{USA}
}
\email{sheng.chen@louisiana.edu}

\begin{abstract}
Reliable learning-based vulnerability detection requires high-quality
labels, yet widely used datasets built from vulnerability-fixing commits
may label functions as vulnerable simply because they were changed by a
security patch. Such label errors can affect both model training and
evaluation. We present \vulvalidate, a framework that uses LLM agents to
coordinate dynamic analysis tools and build vulnerability-triggering
experiments from runtime feedback. Given a labeled function and its fixing
patch, \vulvalidate reconstructs the historical vulnerable and fixed
revisions, analyzes the possible vulnerability, selects suitable tools and
execution paths, iteratively refines triggering inputs, and compares runtime
behavior across the two revisions. The resulting evidence is used to assess
whether the observed defect can be attributed to the labeled function and is
preserved for inspection and replay.
We apply \vulvalidate to all 35,849 instances originally labeled vulnerable
in BigVul, PrimeVul, and DiverseVul. Of these, 20,510 are confirmed
(57.2\%), 6,819 receive label corrections (19.0\%), 7,981 remain attacked
but undecided (22.3\%), and 539 are not successfully measured (1.5\%).
After resolving identified label conflicts and removing byte-exact
duplicates, the corrected release contains 15,890 distinct confirmed
vulnerable function bodies. In a blinded review of 581 decisions randomly
sampled from the three datasets and outcome groups, the final reviewer
consensus supports 90.0\%--92.0\% of sampled confirmations and
92.6\%--99.0\% of sampled label corrections. Keeping model parameters
fixed, evaluation with the corrected labels lowers F1 for all five tested
detectors on both BigVul and DiverseVul. Retraining with corrected labels
then improves F1 over evaluation without retraining for four of five
detectors on each dataset. We also release a reusable \vulvalidate skill
for applying dynamic tools to additional vulnerability instances. The
corrected datasets, reproducible evidence, and reusable skill provide a
stronger basis for vulnerability detection research and future dataset
construction.
\end{abstract}

\maketitle

\section{Introduction}
\label{sec:introduction}

Vulnerability detection is essential to software security, and the growing
volume of reported vulnerabilities increases the need for scalable analysis
and reliable validation. In April 2026, NIST announced that the National
Vulnerability Database would prioritize enrichment according to risk,
following a 263\% increase in CVE submissions between 2020 and
2025~\cite{nist2026nvdofficial}. Learning-based detectors offer a way to
prioritize potentially vulnerable code, but their training and evaluation
depend on the quality of function labels. Inaccurate labels can reward
agreement with dataset artifacts while obscuring whether the underlying
defects are detected~\cite{croft2023dataquality,risse2025wrongexam}.

Widely used datasets, including BigVul, DiverseVul, and PrimeVul, derive
function labels from historical vulnerability information and source
changes~\cite{fan2020bigvul,chen2023diversevul,ding2025primevul}.
Collection rules differ, and PrimeVul refines labelling and
deduplication procedures~\cite{ding2025primevul}. Identifying a vulnerability-fixing
commit and identifying the function responsible for the defect are
different tasks. A repair can change a faulty parser, its callers, tests,
and unrelated code. Labelling changed functions from commit membership
can therefore confuse participation in a repair with responsibility for
the vulnerability. Static-analyzer datasets such as D2A use differential
analysis of warnings across revisions, providing another basis for
candidate labels~\cite{zheng2021d2a}.

Prior audits show that these issues affect benchmark quality. Croft
et al.\ found that 20\% to 71\% of vulnerability labels were inaccurate
in the real-world datasets they examined and that duplication could
inflate model performance~\cite{croft2023dataquality}. Risse et al.\
examined 100 functions labelled vulnerable from each of BigVul, Devign,
and DiverseVul, finding actual vulnerabilities in only 39, 47, and 65
functions, respectively. Their survey also found that 53 of 81
vulnerability-detection papers used at least one of BigVul, Devign, or
ReVeal~\cite{risse2025wrongexam}. Together, these audits motivate
improving the labels used by existing detectors, alongside developing
new architectures.

Our joint analysis identifies conflicting labels across the three studied
datasets. After removing comments and normalizing whitespace, 2,707 of
25,807 unique function bodies labelled vulnerable also appear with a
non-vulnerable label, corresponding to 10.49\%. Of these conflicts, 2,210
occur across datasets. Such matching bodies provide contradictory
supervision when the datasets are combined. Resolving them requires
evidence about the relevant function and source context.

Dynamic analysis provides a useful basis for examining these labels.
Source and patch analysis can identify suspected defects and guide testing,
while execution tests whether a concrete input reaches the relevant
operation and produces an observable failure. A mechanism-appropriate
oracle, together with coverage and attribution evidence, can connect
that failure to the labelled function. Replaying the same condition on
the fixed revision tests whether the repair changes the observed
behavior~\cite{hazimeh2020magma,barr2015oracle,mei2026arvo}.
For example, memory instrumentation can expose an invalid access, while
authorization and resource failures require oracles matched to their
security consequences. Attribution must connect the observed fault to
the labelled function, and a harness must preserve relevant guards and
calling conditions. These requirements help distinguish a defect in the
historical code from a failure introduced by the test
(Section~\ref{sec:oracle}). The scope of the evidence matters as much
as its presence: an unsuccessful search may reflect incomplete coverage,
limited inputs, or an unsuitable oracle, so failure to trigger a defect
is not by itself grounds for a non-vulnerable
label~\cite{klees2018evaluating}.

A major practical obstacle is the cost of obtaining faithful executions
at scale. Commit mining provides candidate labels cheaply, while validating
each function requires mechanism-specific investigation and execution.
Each historical instance may require recovering source revisions,
resolving dependencies, constructing a harness, and finding an input that
reaches the suspected mechanism. Google's OSS-Fuzz documentation notes
that writing fuzz targets typically requires several hours of manual work
and project knowledge. Its account of LLM-based target generation reports
runtime coverage of about 30\% for many integrated
projects~\cite{googleossfuzztargets}. ARVO further shows that rebuilding
historical programs remains difficult even when a triggering input is
already available~\cite{mei2026arvo}. These costs help explain why prior
label audits often examined samples of 50 to 100 functions per
dataset~\cite{chen2023diversevul,croft2023dataquality,risse2025wrongexam}.

Recent LLM agents offer an opportunity to automate much of this work.
SWE-agent and AutoCodeRover combine repository navigation, code analysis,
and tool use to address software
tasks~\cite{yang2024sweagent,zhang2024autocoderover}.
In security testing, Google reported 26 previously unknown vulnerabilities
found through AI-generated fuzz targets in 2024~\cite{google2024aifuzzing}.
DARPA's 2025 AI Cyber Challenge reported that autonomous systems found
54 synthetic vulnerabilities and patched 43 across challenges spanning
54 million lines of code~\cite{darpa2025aixccofficial}. These results
motivate using agents to construct and refine experiments from execution
feedback. Agents can adjust build commands, prepare harnesses, and revise
inputs when an attempt fails to reach the target. Their explanations
still require supporting source and runtime observations before they
can justify a dataset label.

Existing approaches address different parts of this problem. D2A labels
static-analysis findings, and CleanVul uses LLM-based heuristics to identify
vulnerability-related function changes~\cite{zheng2021d2a,li2024cleanvul}.
ARVO automates reproduction of historical vulnerabilities from OSS-Fuzz
and supplies executable environments~\cite{mei2026arvo}. CVE-Factory
already combines agents with dynamic execution to construct vulnerability
tasks and assess source authenticity and repair
validity~\cite{luo2026cvefactory}. VulTrigger uses multiple agents to
generate downstream tests for third-party library
vulnerabilities~\cite{ju7275148vultrigger}. Our focus is auditing existing
function labels in BigVul, PrimeVul, and DiverseVul, then examining
detector performance under the resulting corrections. This requires
connecting runtime observations to the particular labelled function
and defining how confirmation, label correction, and unresolved
attempts follow from the available evidence.

We introduce \vulvalidate, a workflow that uses LLM agents to orchestrate
dynamic analysis tools for reproducible function-level label auditing.
Starting from historical source and the listed fixing commit, the agent
analyzes the mechanism, selects tools and an execution route, constructs
and refines triggering inputs, and performs controlled executions of
vulnerable and fixed revisions. Function execution, oracle relevance,
and causal attribution determine what the observations support.
Label correction requires an attack satisfying seven evidence checks and
a source-grounded explanation rejecting the original attribution under
the tested conditions (Section~\ref{sec:negative}). Source bindings,
inputs, logs, return codes, coverage, and attack histories support replay
and inspection. Instances with insufficient evidence retain an explicit
unresolved outcome. The workflow separates defect existence, correspondence
to the listed CVE, and repair effectiveness, with function-level attribution
required to assess the dataset label.

We apply \vulvalidate to all 35,849 instances originally labelled
vulnerable in BigVul, PrimeVul, and DiverseVul. The audit yields 20,510
confirmations, 6,819 label-correction decisions, 7,981 attacked but
undecided instances, and 539 instances that were not successfully
measured. Two of the authors, each with vulnerability-detection experience, blinded to the agent's
decisions, each reviewed the same 581 instances, randomly sampled within
each dataset and initial S6 outcome group; their consensus supports
90.0\% to 92.0\% of sampled confirmations and 92.6\% to 99.0\% of
sampled label corrections (Section~\ref{sec:review-results}), and agreement with independently produced manual labels yields
Cohen's $\kappa$ values~\cite{cohen1960kappa} from 0.36 to 0.55, providing an external
reference for the audit decisions.
(Section~\ref{sec:external-results}). The corrected datasets retain
confirmed and corrected instances whose decisions survive review, and
exclude unresolved instances and unresolved label conflicts
(Section~\ref{sec:dataset-construction}).

The model experiments distinguish re-evaluation from retraining.
With model parameters held fixed, corrected evaluation protocols lower F1
for all five detectors on BigVul and DiverseVul, with absolute decreases
of 0.0403 to 0.2737 and 0.0532 to 0.0888, respectively.
On corrected test labels, the completed retraining runs achieve higher
F1 for four of five detectors on each dataset
(Section~\ref{sec:model-results}). The merged benchmark further provides
a post-correction evaluation of graph-based and transformer-based
detectors, connecting label auditing to the interpretation of model
performance.

The study addresses one design question and three empirical questions:
\begin{description}[leftmargin=3em,style=nextline,itemsep=2pt]
\item[\ourrq{1}] How do LLM agents construct reproducible dynamic
experiments to audit function-level vulnerability labels?
\item[\ourrq{2}] What are the results of manually reviewing the agents'
audit decisions?
\item[\ourrq{3}] What are the distributions of audit outcomes
across BigVul, PrimeVul, and DiverseVul?
\item[\ourrq{4}] What changes in vulnerability detector performance
are observed after label correction?
\end{description}

This paper makes three contributions.
\begin{enumerate}[leftmargin=*,itemsep=2pt]
\item \textbf{Improving label quality for three widely used datasets.}
We provide an audit of the complete populations originally labelled
vulnerable in BigVul, PrimeVul, and DiverseVul, together with corrected
data constructed through label adjudication, manual-review filtering,
conflict resolution, and deduplication. Original labels, audit outcomes, supporting evidence, and
unresolved instances remain available for inspection. This supports reuse
of the corrected data while preserving the conditions under which each
label was assessed.

\item \textbf{Agentic dynamic validation of historical vulnerability labels.}
\vulvalidate combines historical reconstruction, feedback-guided trigger
construction, controlled execution, and function-level attribution under
an explicit evidence contract. The operational prompt, tool and oracle
rules, and evidence requirements form a reusable skill for auditing
additional vulnerability instances.

\item \textbf{A reproducible vulnerability evidence corpus.}
The release contains 15,890 distinct confirmed vulnerable
function bodies after byte-exact deduplication, with retained source
versions, triggering inputs or harnesses, execution logs, oracle
observations, attribution evidence, and replay materials. This resource supports vulnerability detection,
localization, repair evaluation, and further label review.
\end{enumerate}

\begin{figure*}[t]
    \centering
    \includegraphics[width=\textwidth]{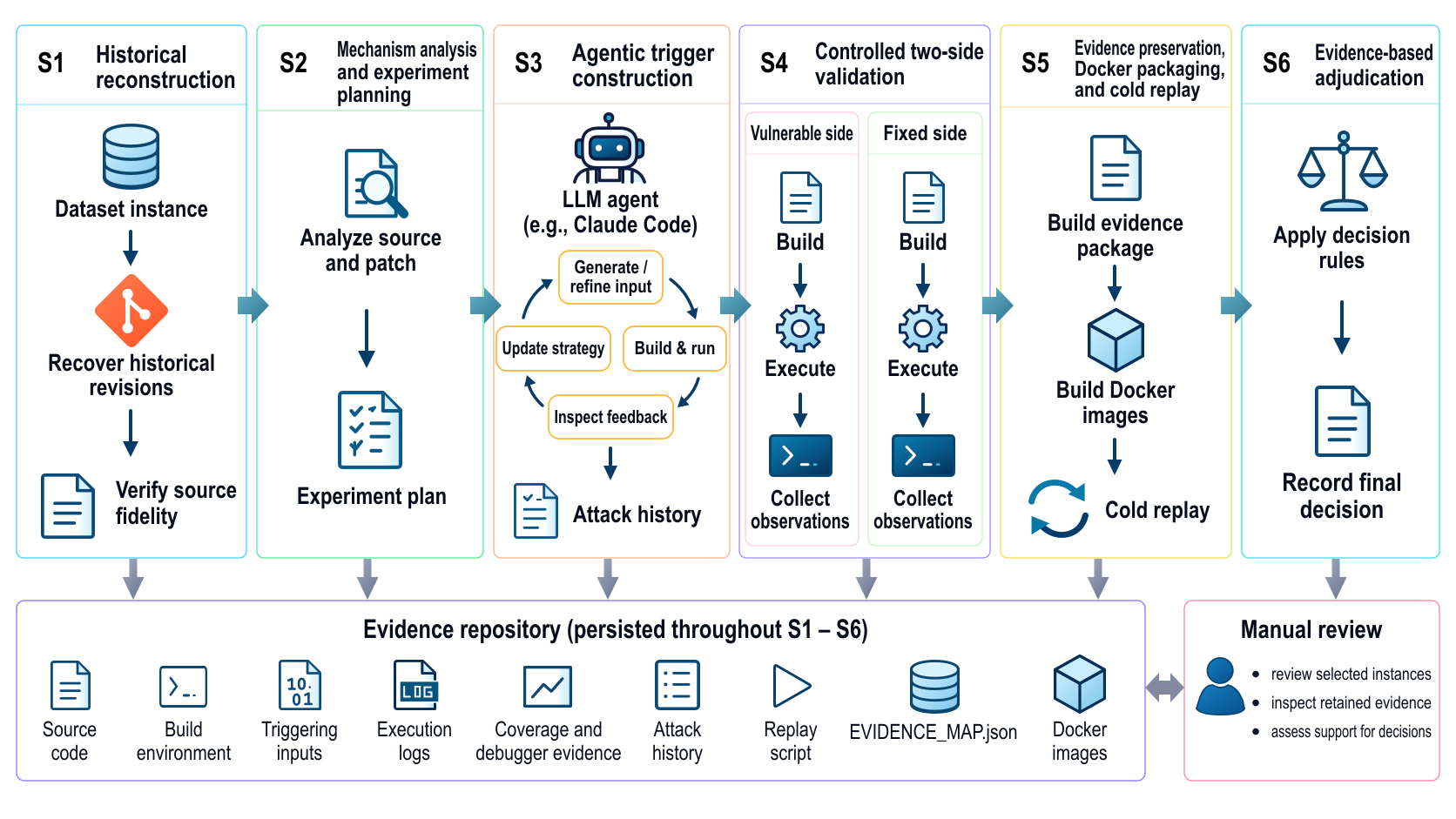}
    \caption{Overview of \vulvalidate. The workflow reconstructs historical
    source and environment (S1), plans the experiment (S2), iteratively
    constructs triggers (S3), validates vulnerable and fixed executions (S4),
    preserves evidence and Docker-based replay materials (S5), and assigns an
    evidence-based audit outcome (S6). Two experts review randomly sampled
    instances blinded to the agent's decisions.}
    \label{fig:pipeline}
\end{figure*}

\section{The \vulvalidate Method}
\label{sec:method}

\vulvalidate audits a labelled function using historical source,
vulnerability metadata, and a listed fixing commit.
Figure~\ref{fig:pipeline} summarizes six stages: reconstruction
and planning (S1--S2), feedback-guided trigger construction (S3),
controlled execution and attribution (S4), evidence preservation
and cold replay (S5), and adjudication (S6).
Retained source and runtime evidence determine whether the
original attribution is supported, rejected within the tested
mechanism, or unresolved.

\subsection{Historical reconstruction and experiment planning (S1--S2)}
\label{sec:controlled}

S1 reconstructs a historical testbed and S2 specifies how to exercise it,
following the reconstruction requirements that ARVO and CVE-Factory
established for executable vulnerability
reproduction~\cite{mei2026arvo,luo2026cvefactory}.

\parag{S1: Reconstruct the historical target and execution environment.}
Using the listed fixing commit, \vulvalidate locates the upstream
repository, verifies the commit against repository history and the
vulnerability metadata, and takes its parent as the initial vulnerable
revision. Let $f_D$ denote the dataset function body, $f_v$ the
corresponding function at the verified vulnerable revision, and $f_f$ its
counterpart at the fixed revision. The label attached to $f_D$ is audited
on $f_v$ and $f_f$; the repository, exact commits, file and function
extent, and fixing diff are recorded as source bindings. Environment
reconstruction follows standard practice for historical builds: the
source tree, dependencies, toolchain, and platform are recovered from the
repository and its metadata, in Docker containers where the user-space
environment can be reproduced directly and in virtual machines where
kernel or platform requirements need stronger
isolation~\cite{merkel2014docker,bellard2005qemu,kivity2007kvm}.

Source fidelity is checked separately from build success, because a
dataset body that compiles may still omit guards, state updates, or calls
relevant to the suspected mechanism. \vulvalidate compares $f_D$ with
$f_v$ for identity, extent, and completeness; if $f_D$ is incomplete, the
experiment uses the verified historical function and records the
discrepancy.

\parag{S2: Analyze the vulnerability mechanism and plan the experiment.}
The agent analyzes the fixing diff, $f_v$, $f_f$, callers and callees,
and the CVE and CWE metadata, with verified source as the primary basis
when metadata is missing or inconsistent. The plan records a root-cause
hypothesis, the attacker- or environment-controlled precondition, the
disputed operation or branch, the entry point, the execution route, and
the tools and oracle selected for the mechanism (Section~\ref{sec:agent});
Section~\ref{sec:example} shows one such plan.

Two execution routes are available. The full-project route builds the
historical project and exercises a public interface that reaches the
labelled function in its real program
context~\cite{mei2026arvo,google2026ossfuzz}. When this is infeasible,
the extracted-function route compiles the real target with its
mechanism-relevant types, guards, callees, and calling conditions behind
a harness; stubs may replace unavailable services only while preserving
these conditions. The route fixes the evidence scope: full-project
execution supports reachability and comparison across revisions, but when
a fixing commit changes several functions, a difference between revisions
establishes repair-associated behavior and attribution to the labelled
function requires further evidence; extracted-function execution isolates
the target by varying it while holding harness, input, and context fixed.
The route and reconstructed context are recorded for S4.

\subsection{Agentic trigger construction (S3)}
\label{sec:agent}

S3 repeatedly executes attack attempts against the vulnerable side using
the S2 plan, historical source, dataset metadata, fixing diff, prior attack
history, and build information. Runtime observations guide changes to
inputs, harnesses, execution state, tools, or routes.

The study uses Claude Code as the execution
controller~\cite{anthropic2026claudecode,anthropic2026opus5}, with model and
thinking-effort settings in Section~\ref{sec:agent-config}. It navigates
repositories, edits files, executes commands, and operates analysis tools,
following capabilities demonstrated by repository
agents~\cite{yang2024sweagent,zhang2024autocoderover,jimenez2024swebench}.
CVE-Factory constructs executable vulnerability tasks, VulTrigger generates
downstream tests, and OSS-Fuzz agents operate in build and test
environments~\cite{luo2026cvefactory,ju7275148vultrigger,zhang2026coderover}.
An operational prompt defines source rules, route preferences, tool and
oracle requirements, evidence layout, outcome criteria, and stopping
conditions. The prompt constrains the evidence rather than the controller,
so another model can run the same loop against the same requirements.

Tool selection follows the suspected mechanism and relies on standard
dynamic analysis tools: sanitizers for memory, uninitialized-value,
arithmetic, leak, and concurrency errors (ASan/KASAN, MSan/KMSAN, UBSan,
LSan, TSan/KCSAN, and
Memcheck)~\cite{serebryany2012asan,linux2026kasan,stepanov2015msan,linux2026kmsan,llvm2026ubsan,llvm2026lsan,serebryany2009tsan,linux2026kcsan,nethercote2007valgrind};
progress, termination, and resource observations for denial of service;
security-decision, state, or output oracles for authorization, injection,
and state-machine defects~\cite{barr2015oracle}; and coverage and debugger
traces for execution and attribution evidence~\cite{hazimeh2020magma}.

The controller runs a feedback-guided loop in the style of greybox and
sanitizer-guided
fuzzing~\cite{fioraldi2020aflpp,bohme2017aflgo,osterlund2020parmesan,meng2024chatafl}.
Each iteration instantiates the S2 hypothesis with a route, entry point,
tools, oracle, and required state; constructs or modifies an input,
schedule, or harness while preserving the calling conditions identified
in S2; builds the pinned vulnerable environment with the selected
instrumentation and executes it under documented resource limits; and
collects oracle output, coverage, traces, and return status. Following
Magma~\cite{hazimeh2020magma}, the observations are read at three levels:
function entry establishes reachability, coverage or tracing establishes
execution of the disputed operation or branch, and the oracle establishes
a security-relevant violation. The gap determines the next refinement: an
unreached function revises the input or entry path, an unexecuted
disputed operation revises the triggering condition, harness, or state,
and an inadequate observation revises the tools or oracle. When a trigger
succeeds, the exact input, harness, source and environment bindings,
commands, logs, and coverage are frozen for two-side replay in S4;
unsuccessful attempts are kept with their conditions, scope, and stopping
reason, since they document the search scope that check~5 of
Table~\ref{tab:negative} requires. The loop ends when sufficient evidence
is obtained or a predefined stopping condition is reached. Tool selection
rules, the full step list, and stopping conditions are given in \supp{2}.

\subsection{Controlled two-side validation and function-level attribution (S4)}
\label{sec:oracle}

S4 replays the triggering condition retained by S3 on the vulnerable and
fixed revisions to test whether the listed repair changes the observed
security-relevant behavior~\cite{mei2026arvo,luo2026cvefactory}.
Let $d$ denote the driver or public-entry procedure, $x$ the saved input
or schedule, $e$ the environment, and $O$ the mechanism-specific oracle:
\begin{equation}
 y_v = O(d[f_v],x,e), \qquad
 y_f = O(d[f_f],x,e).
 \label{eq:experiment}
\end{equation}
The executions share the driver, input, relevant build and instrumentation
settings, runtime environment, resource limits, timeout policy, and oracle
settings. The intended variable is the pinned source revision. Both
binaries are hashed after construction to check whether the builds are
distinct. Matching binaries are documented as build-equivalent outcomes.
When an interface change prevents identical inputs, the exception is
documented and the comparison is narrowed accordingly.

The mechanism-specific oracle selected in S2 and applied in S3 interprets
sanitizer reports, output or state changes, progress, or other observations
relevant to the suspected mechanism~\cite{barr2015oracle}.
Crash and exit-code comparisons are used when appropriate, with the
security-relevant mechanism or state determining the interpretation.
For function-level attribution, the evidence connects
\[
\text{input}
\rightarrow
\text{target reached}
\rightarrow
\text{mechanism triggered}
\rightarrow
\text{violation observed}.
\]
Function entry establishes reachability. Attribution additionally requires
execution of the relevant operation, branch, or state transition and its
causal connection to the violation~\cite{hazimeh2020magma}. A fault in a
callee requires evidence linking the labelled function's action to the
affected object or state and the downstream observation. Both the comparison
and attribution evidence are retained for S5 and S6.

\subsection{Evidence preservation, Docker packaging, and cold replay (S5)}
\label{sec:evidence}

S5 organizes the evidence and attack history collected throughout
the workflow into replayable packages. Container-compatible experiments
are distributed as versioned Docker images containing the source code,
execution environment, replay scripts, and supporting evidence.
Cold replay checks whether the experiment can be reconstructed and
executed from the retained materials. ARVO distributes Docker images containing historical source,
build environments, and triggering inputs for recompilation and
replay~\cite{mei2026arvo}. CVE-Factory likewise bases evaluation on
executable artifacts~\cite{luo2026cvefactory}. \vulvalidate additionally
preserves the function-level evidence needed for label adjudication.

The package contains source bodies and revision identifiers, the driver
or harness, dependency and build information, unchanged inputs or saved
schedules, available vulnerable and fixed logs, actual integer return
codes, execution and attribution evidence, and the oracle rationale.
A replay script specifies reconstruction and execution, and
\path{EVIDENCE_MAP.json} lists each active file's purpose, path, size,
and SHA-256 digest. Attack histories retain tested conditions, routes,
tool settings, observations, coverage, limitations, refinement reasons,
and stopping conditions. Failed attempts are kept alongside successful
ones; they document the search scope that check~5 of
Table~\ref{tab:negative} requires.

To support reproduction, the container-executable experiments are
released as Docker images. The images bundle
recovered source code, compilers, dependencies, build and replay scripts,
and the complete evidence package, including attack histories.
Vulnerable and fixed builds are kept separate, with documented commands
for rebuilding and replaying each side. Dockerfiles and pinned base-image
references accompany the images. Following S6 adjudication, a release
index links instance identifiers and audit outcomes to image digests
and replay commands.

Before an experiment is treated as reproducible, the workflow attempts a
cold replay. For containerized experiments, replay runs in a fresh
container without access to the original workspace. Generated build
products are removed before rebuilding from the retained source and
inputs. Original evidence remains unchanged, with new replay outputs
stored separately for comparison. Source and binary identities, oracle
observations, and evidence-file hashes are checked. Build information
identifies the repository, revision, compiler, dependencies, settings,
and local modifications. Experiments requiring specific kernels,
devices, or virtual machines document these additional requirements,
consistent with S1.

A complete package does not by itself establish a label: manual review
(Section~\ref{sec:review-design}) examines the claimed defect and
function attribution, the harness's calling conditions, and the source
reasoning behind a label correction. The supplementary artifact
provides the prompt or skill, field definitions, route and oracle rules,
package layout, stopping conditions, and integrity requirements.

\subsection{Evidence-based interpretation and adjudication (S6)}
\label{sec:interpretation}

S6 assigns four outcomes from the retained evidence. A qualifying
target-attributed witness supports confirmation; otherwise, a supported
rejection of the original attribution supports label correction.
Unresolved outcomes depend on whether a usable dynamic observation exists.

\parag{Confirmed.}
Confirmation requires executing the complete, source-bound historical
function on the dataset-labelled side under a saved input or runtime
condition. The driver preserves relevant types, guards, and calling
contracts. A mechanism-appropriate oracle must identify a real defect or
security-relevant violation causally linked to the function
(Section~\ref{sec:oracle}). A valid witness supports confirmation despite
a CVE mismatch or persistence on the comparison side. Source authenticity,
dynamic execution, and repair validity remain separate
claims~\cite{luo2026cvefactory}.

Controlled failure injection requires an upstream repair explicitly
addressing the injected condition, injection through a public interface
preserving upstream code, and the same injector on both sides.
Evidence must link the condition to the fixing diff, show it occurred
and reached the disputed operation, and retain a working positive control.
Attacker-controlled reachability requires separate evidence.

\parag{Label correction.}
\label{sec:negative}
Label correction rejects the original attribution within the tested
mechanism and conditions. Because an unsuccessful search depends on
inputs, coverage, budget, and oracle~\cite{klees2018evaluating}, a
qualifying dynamic negative requires an attack yielding no
target-attributed defect, all seven checks in Table~\ref{tab:negative},
and a source-grounded rejection rationale. The attack must exercise the
condition addressed by that rationale. Static observations guide analysis;
search exhaustion or equal behavior alone leaves attribution unresolved.

\begin{table}[t]
\caption{Seven necessary checks for a mechanism-specific dynamic
negative. Each requires inspectable source or execution evidence.}
\label{tab:negative}
\tablebody
\begin{tabularx}{\linewidth}{@{}rL{0.23\linewidth}Y@{}}
\toprule
 & Check & Required evidence \\
\midrule
1 & Measured execution &
A replay script, actual integer exit codes, and complete logs,
including both sides for a two-sided claim \\

2 & Function execution &
Trace, coverage, or debugger evidence of labelled-function execution \\

3 & Attack input &
A saved input, failure condition, or schedule targeting the mechanism,
with construction and purpose documented \\

4 & Oracle relevance &
An oracle capable of observing the suspected defect and a
mechanism-specific rationale~\cite{barr2015oracle} \\

5 & Search scope &
Tested entry points, value ranges, corpus or schedules, duration or
iteration bounds, and stopping reason~\cite{klees2018evaluating} \\

6 & Source completeness &
Comparison of the tested instance with the complete function at the
verified upstream parent revision \\

7 & Patch execution &
Hit counts or equivalent runtime evidence that the disputed operation,
line, or branch executed~\cite{hazimeh2020magma} \\
\bottomrule
\end{tabularx}
\end{table}

Let $C_1,\ldots,C_7$ denote these checks and $R$ the evidence-supported
rejection rationale. The acceptance condition is
\begin{equation}
    N_{\mathrm{corr}}
    =
    \left(\bigwedge_{j=1}^{7} C_j\right)
    \land R.
    \label{eq:negative-contract}
\end{equation}
The rationale explains why the exercised behavior rejects the attribution,
for example an existing guard correctly handling the challenged condition.
An apparent inverted differential, with a defect observed only on the
comparison side, requires verification of source orientation, source and
build bindings, and the observed defect, and still needs the rejection
rationale $R$; otherwise the instance remains
\emph{Attacked but undecided}, because the comparison-side witness
concerns only that side's code.

\parag{Attacked but undecided.}
An executed attack yields observations but insufficient evidence for
confirmation or label correction, including unresolved fault signals.
The instance retains observations, tested scope, evidence gaps, and
stopping reason. A two-sided claim requires runtime evidence from both
executions; a single-side execution alone leaves the instance here.

\parag{Not successfully measured.}
This outcome applies when the workflow ends before obtaining a usable
dynamic observation. Barriers include historical source recovery,
dependencies, platform requirements, toolchain problems, or execution
routes, with build failure being one possible cause~\cite{mei2026arvo}.

\subsection{An example: one commit, four labels, one defect}
\label{sec:example}


Commit \texttt{3e5cdf1} of \texttt{facebook/zstd} is the listed fix for
CVE-2019-11922,\footnote{%
\url{https://nvd.nist.gov/vuln/detail/CVE-2019-11922};
\url{https://github.com/facebook/zstd/commit/3e5cdf1b6a85843e991d7d10f6a2567c15580da0}.}
reported as a race condition that ``could allow an attacker to write bytes
out of bounds if an output buffer smaller than the recommended size was
used'' and assigned CWE-362. It changes four compressor functions, all four
of which BigVul labels vulnerable, as instances 182843--182846. One hunk
changes run-time behavior; the other three add \texttt{DEBUGLOG} calls.

\parag{S1.}
The parent revision \texttt{3a90229} supplies $f_v$, which the dataset body
matches except for leading whitespace on ten lines. The testbed is the whole
library at that revision, with 39 of its 40 files byte-identical to upstream
by SHA-256 and the 40th substituting the stored body under test, so it
compiles in its real translation unit.

\parag{S2.}
From the diff the agent recorded this plan.
\begin{lstlisting}[caption={Experiment plan recorded for BigVul 182843, abridged.},captionpos=b]
root cause  : the set_rle arm stores *op = codeTable[0] before any test of
              dstCapacity, and upstream's only check is taken before the
              sequences header, so a later table can be entered with
              dstCapacity == 0
precondition: the caller supplies dst smaller than ZSTD_compressBound()
entry point : the public one-pass API ZSTD_compress()
oracle      : an ASan heap-buffer-overflow WRITE with frame #0 inside
              ZSTD_buildCTable -- the byte lands in whatever follows dst,
              so without a sanitizer nothing crashes or returns an error
\end{lstlisting}

\parag{S3.}
Three rounds are retained. The first compiled a harness whose own body
performed the store; ASan fired, but its frame \#0 lay in the harness file,
no library code ran, and there was no fixed side, so the record marks the
attempt superseded and keeps it. The second wrote an end mark past each
truncated output capacity, across levels 1--19 and 40 input shapes, to find
a capacity at which the compressed offset table ends exactly on the buffer's
last byte. The third fixed the input to 40 records whose equal match lengths
make upstream choose \texttt{set\_rle} at that capacity, so the match-length
table is entered with \texttt{dstCapacity == 0}.

\parag{S4.}
Replayed on both revisions with the same driver, input and limits, the
vulnerable side exits 1 with an ASan WRITE of size one ``0 bytes to the
right'' of the 192-byte destination while the fixed side exits 0 returning
\texttt{dstSize\_tooSmall}. Frame \#0 identifies the disputed store in
\texttt{ZSTD\_buildCTable}, and frames \#1--\#10 show the executed
call path from \texttt{ZSTD\_compress}. Adding the capacity check
in \texttt{ZSTD\_buildCTable} removes the observed write.
Following Section~\ref{sec:oracle}, function-level attribution
depends on mechanism-specific source and runtime evidence;
a caller's presence in the stack alone is insufficient for
attribution. Table~\ref{tab:example} summarizes the four decisions.

\begin{table}[t]
\caption{The four functions labelled vulnerable by the single fixing commit
\texttt{3e5cdf1}, and the evidence that decided each.}
\label{tab:example}
\tablebody
\small
\begin{tabularx}{\linewidth}{@{}L{0.25\linewidth}L{0.08\linewidth}L{0.13\linewidth}Y@{}}
\toprule
Labelled function & BigVul index & Role in the fault & Outcome and deciding evidence \\
\midrule
\texttt{ZSTD\_buildCTable} &
182843 &
frame \#0 &
\emph{Confirmed.} ASan WRITE of size 1 past a 192-byte buffer at the
disputed store; exit codes 1 against 0 \\
\toprule
\texttt{ZSTD\_compress\-Sequences\_internal} &
182844 &
frame \#1, a caller &
\emph{Label correction.} gcov: 12 calls per side, disputed line 3 times per
side, no fault \\
\toprule
\texttt{ZSTD\_encodeSequences} and \texttt{\ldots\_body} &
182845, 182846 &
absent from the stack &
\emph{Label correction.} 3 calls per side, disputed lines executed on both
sides, no fault \\
\bottomrule
\end{tabularx}
\end{table}

\parag{S5--S6.}
Replayed in a fresh container with no network and no access to the original
workspace, the instance gives the same exit codes, frame and return value. It
is \emph{Confirmed}. The assigned CWE-362 does not describe the observed
mechanism, a missing capacity check rather than a race; per
Section~\ref{sec:interpretation} a valid witness still supports confirmation
and the record keeps the disagreement. The three siblings are assigned label
correction under Equation~\eqref{eq:negative-contract}, each satisfying the
seven checks of Table~\ref{tab:negative}. \supp{5} gives the attack history,
the full report and call chain, the per-sibling evidence, and the replay
transcript.

\parag{Answer to \ourrq{1}.}
LLM agents construct reproducible experiments by linking historical
reconstruction and planning to feedback-guided attacks, controlled
two-side execution, and function-level attribution. Each decision is
bound to retained source, inputs, logs, and coverage that replay in a
fresh container, and the S6 contract, not the agent's explanation,
decides the outcome.

\section{Experimental Design}
\label{sec:design}

\subsection{Audit population and agent setup}
\label{sec:frames}
\label{sec:agent-config}

The audit covers 35,849 instances originally labelled vulnerable:
10,900 from BigVul, 6,004 from PrimeVul, and 18,945 from the official
DiverseVul collection~\cite{fan2020bigvul,ding2025primevul,chen2023diversevul}.
Each instance is a labelled function and receives one of the four S6
outcomes (Section~\ref{sec:interpretation}); reported outcome
distributions incorporate the manual-review results, and the initial
agent decisions are retained for the review comparison
(Section~\ref{sec:review-design}). The agent was Claude Opus~5.0 run
through Claude Code with thinking effort set to \texttt{high} throughout
S1 to S6; it reconstructed programs, executed attacks, preserved
evidence, and assigned outcomes under the S6 criteria.

\subsection{Manual review}
\label{sec:review-design}
\parag{Blinded two-expert review.}
Two of the authors, each with vulnerability-detection experience, each reviewed all 581 selected
instances using a common checklist. The agent's decisions were withheld
throughout individual assessment and consensus discussion. Reviewers
examined source code, inputs, harnesses, runtime logs, and attribution
evidence to form their own judgments under the S6 criteria. Each
instance received two individual assessments, and disagreements were
resolved through discussion to obtain one final consensus assessment.

For all three datasets, instances were selected by simple random sampling
without replacement within each dataset and initial S6 outcome group.
The sample contains 300 instances initially assigned to \emph{Confirmed}
and 281 assigned to \emph{Label correction }. BigVul and
DiverseVul each contribute 100 instances per group; PrimeVul contributes
100 confirmations and 81 label-correction instances. Sampling used the
full pool within each group. Outcome groups define sampling and
reporting, not the information shown to reviewers.

The checklist follows CVE\mbox{-}Factory's source authenticity, dynamic
execution, and repair validity dimensions~\cite{luo2026cvefactory} and
S6. It covers source completeness, trigger fidelity, execution of the
labelled function and disputed operation, oracle relevance, and causal
attribution. Assessing a possible label correction also requires a
source-grounded rejection rationale supported by dynamic evidence
under the tested conditions (Section~\ref{sec:negative}).

After consensus was finalized, the expert assessments were compared
with the initial agent decisions to obtain \emph{Supported},
\emph{Contradicted}, or \emph{Insufficient evidence}. These verdicts
indicate whether the evidence supports the agent decision, conflicts
with it, or lacks essential support within its scope.

Support rates are $S/n$ by dataset and initial agent outcome, where $S$
counts supported decisions and $n$ includes all reviewed instances,
each counted once, including those subsequently excluded. The rates
describe the sampled outcome groups, not label accuracy across the full
datasets.

\parag{Comparison with independent manual labels.}
\label{sec:external-design}
The review above used this project's reviewers and criteria. As an
external check, we compare the agent's decisions with manual labels
produced independently by two prior studies. Risse
et al.~\cite{risse2025wrongexam} relabelled 100 BigVul and 100 DiverseVul
functions, judging whether each function is vulnerable. Croft
et al.~\cite{croft2023dataquality} checked 70 BigVul functions, all among
Risse et al.'s 100, judging whether each label is accurate. BigVul functions are joined by
row number, which is also our instance identifier, and DiverseVul
functions by index and function-text hash; all 270 join. \emph{Confirmed}
counts as the label standing and \emph{Label correction } as
rejected; the two unresolved outcomes are not decisions. Each agent
decision is read from the campaign's measurement ledger, which records
per function the outcome this campaign's own runs established and how
deep the supporting evidence is, so all four comparisons are on one
scale (\supp{4}). We report raw
agreement, chance agreement from the marginal rates, and Cohen's
$\kappa$~\cite{cohen1960kappa}; the agreement between the two human
studies on their 70 shared functions is the reference. Details and every
disagreement are in \supp{4}.

\subsection{Construction of corrected datasets}
\label{sec:dataset-construction}

Dataset construction uses the outcomes finalized after incorporating
manual review. Among instances originally labelled vulnerable,
\emph{Confirmed} maps to vulnerable ($y=1$), and \emph{Label correction} maps to non-vulnerable ($y=0$) as an evidence-supported empirical label. \emph{Attacked but undecided} and
\emph{Not successfully measured} remain unresolved and are excluded
from both learning classes.

Among the 581 reviewed instances, 545 received a \emph{Supported}
verdict, 12 were \emph{Contradicted}, and 24 had
\emph{Insufficient evidence}. The 545 supported instances pass the
manual-review filter. The remaining 36 instances are excluded from
both learning classes and are conservatively grouped with
\emph{Attacked but undecided} in the post-review audit distribution.
Their initial agent decisions and expert verdicts remain available
in the released audit records. Conflict resolution and deduplication
are applied subsequently.

For label-focused detector comparisons, retained instances keep the
original dataset representation $f_D$. Audit decisions are transferred
only when the source content relevant to the decision is preserved
between $f_D$ and $f_v$; instances with substantive source discrepancies
are excluded from corrected learning and evaluation.

Non-conflicting original non-vulnerable labels are retained. Dynamic
execution and function-level attribution determine the retained label
for conflicting copies. Rejected-label copies and unresolved conflicts
are excluded. Duplicate bodies are consolidated under the applicable
rule.

These rules define the final corrected BigVul, PrimeVul, and DiverseVul
datasets and their merged benchmark. Original-label experiments retain
the original releases. The \emph{Distinct total} uses SHA-256
hashes of unmodified UTF-8 function bodies, preserving comments and
whitespace. The Introduction's conflict analysis instead removes comments
and normalizes whitespace. Model-specific deduplication keys, label sets,
and supported inputs are described in Sections~\ref{sec:model-design}
and~\ref{sec:combined-results}.

Relabelling, exclusion of unresolved and unsupported instances,
deduplication, and conflict resolution change class proportions and
sample membership between the original and corrected benchmarks.

\subsection{Vulnerability detectors}
\label{sec:detectors}

We evaluate three graph-based and two transformer-based detectors.
Devign uses a gated graph neural network over syntax, control-flow,
data-flow, and token-order edges~\cite{zhou2019devign}. ReVeal combines
a similar encoder with representation learning and class
rebalancing~\cite{chakraborty2022reveal}. DeepDFA embeds abstract dataflow
facts over control-flow graphs~\cite{steenhoek2024deepdfa}. LineVul
fine-tunes encoder-only CodeBERT on function
tokens~\cite{fu2022linevul,feng2020codebert}; CodeT5 fine-tunes a pretrained
encoder--decoder through its defect-detection head~\cite{wang2021codet5}.
Details appear in \supp{6}.

\subsection{Vulnerability detector performance evaluation protocol}
\label{sec:model-design}

Let $O$ and $C$ denote the original and corrected dataset settings. In
$X/Y$, $X$ specifies the training data and labels, and $Y$ the test data
and labels. $O/O$ is the
original-label baseline, $O/C$ re-evaluates the $O/O$-trained model against
corrected labels with prediction scores fixed, and $C/C$ uses corrected
labels for training and evaluation. Comparing $O/C$ with $C/C$ assesses
retraining. We omit $C/O$ because original label errors can penalize
predictions consistent with corrected labels.
Let $E$ denote a common test set for each detector and dataset. For BigVul and DiverseVul, $E$ removes only \emph{Label correction} instances from the corrected test set, leaving all remaining
instances and labels unchanged.
Confirmed instances remain positive ($y=1$) and retained originally
non-vulnerable instances remain negative ($y=0$); attacked-but-undecided
and unsuccessfully measured instances remain excluded.
For each detector, $O/E$ and $C/E$ compare the original-trained and
corrected-trained models on the same test instances without treating
label-correction instances as test negatives.

Thresholds are selected on the corresponding validation data, with
established fixed thresholds retained where applicable. The $O/O$ and $C/C$
validation labels are original and corrected, respectively. On BigVul,
LineVul uses 0.5; $O/C$ retains CodeT5's original threshold and reselects
thresholds for DeepDFA, Devign, and ReVeal using corrected validation
labels. DiverseVul retains all $O/O$ thresholds in $O/C$. The merged
benchmark uses validation-selected thresholds, except that LineVul uses
0.5.

We report accuracy, precision, recall, F1, Matthews correlation coefficient
(MCC)~\cite{matthews1975comparison}, and area under the ROC curve
(AUROC)~\cite{fawcett2006introduction}. Under class imbalance, accuracy
is read alongside precision, recall, F1, and MCC, which accounts for all
four confusion-matrix cells~\cite{chicco2020advantages,saito2015precision,yao2020assessing}.
Evaluation populations follow the construction policy and detector input
requirements (Section~\ref{sec:dataset-construction}).

\begin{table}[t]
\captionsetup[subtable]{justification=centering,singlelinecheck=false,skip=2pt}
\caption{Validation of the agent's decisions. In (a), Risse and Croft
denote the manual labellings of Risse et al.~\cite{risse2025wrongexam}
and Croft et al.~\cite{croft2023dataquality}, $n$ counts functions with a
decision on both sides, and chance is the agreement expected from the
marginal rates. In (b), S, C, and I denote Supported, Contradicted, and
Insufficient evidence; $S/n$ includes all reviewed instances, including
those later excluded from the final datasets; Label correction denotes the S6 outcome defined in
Section~\ref{sec:negative}. Agreement,
chance, and $S/n$ are percentages.}
\label{tab:validation}
\begin{subtable}[t]{0.45\linewidth}
\caption{Agreement with independent manual labels}
\label{tab:external}
\tablebody\footnotesize\setlength{\tabcolsep}{3pt}
\centering
\begin{tabular}{@{}llrrrr@{}}
\toprule
Dataset & Comparison & $n$ & Agreement & Chance & $\kappa$ \\
\midrule
\multirow{3}{*}{BigVul} & Agent--Risse & 92 & 77.2 & 49.5 & 0.55 \\
                        & Agent--Croft & 63 & 68.2 & 50.1 & 0.36 \\
                        & Risse--Croft & 70 & 77.1 & 48.9 & 0.55 \\
\midrule
DiverseVul              & Agent--Risse & 87 & 78.2 & 55.9 & 0.50 \\
\bottomrule
\end{tabular}
\end{subtable}\hfill%
\begin{subtable}[t]{0.53\linewidth}
\caption{Blinded expert review of 581 sampled decisions}
\label{tab:review}
\tablebody\footnotesize\setlength{\tabcolsep}{2.5pt}
\centering
\begin{tabular}{@{}lrrrrrrrrrr@{}}
\toprule
\multirow{2}{*}{Dataset} &
\multicolumn{5}{c}{Confirmed} &
\multicolumn{5}{c@{}}{Label correction} \\
\cmidrule(lr){2-6} \cmidrule(l){7-11}
 & $n$ & S & C & I & $S/n$ & $n$ & S & C & I & $S/n$ \\
\midrule
BigVul     & 100 & 92 & 3 & 5  & 92.0 & 100 & 99 & 0 & 1 & 99.0 \\
PrimeVul   & 100 & 91 & 6 & 3  & 91.0 & 81  & 75 & 2 & 4 & 92.6 \\
DiverseVul & 100 & 90 & 0 & 10 & 90.0 & 100 & 98 & 1 & 1 & 98.0 \\
\bottomrule
\end{tabular}
\end{subtable}
\end{table}

\section{Experimental Results}
\label{sec:results}

\subsection{External validity: agreement with independent manual labels}
\label{sec:external-results}

Table~\ref{tab:external} reports agreement on the functions where both
sides reached a decision: 92 of Risse et al.'s 100 BigVul functions (48
\emph{Confirmed}, 44 \emph{Label correction}), 63 of Croft et al.'s 70,
and 87 of Risse et al.'s 100 DiverseVul functions (59 and 28). The
remainder are \emph{Attacked but undecided} or \emph{Not successfully
measured} and are neither agreement nor disagreement.

The agent agrees with Risse et al.\ on 71 of 92 BigVul functions
($\kappa=0.55$), as closely as the two human studies agree with each
other ($\kappa=0.55$), and on 68 of 87 DiverseVul functions
($\kappa=0.50$): moderate agreement on the scale of Landis and
Koch~\cite{landis1977kappa}. Seventeen of the 21 disagreements are confirmations that Risse et al.\
call secure. For seven of them, Risse et al.\ attribute the reported
CVE's defect to another function in the same commit; for the other ten,
they judge the commit not to be a security fix. Under
Section~\ref{sec:interpretation}, confirmation requires dynamic
evidence of a defect or security-relevant violation causally
attributable to the labelled function. Correspondence to the reported
CVE and the effectiveness of the listed repair are assessed separately.
These cases remain documented disagreements, with instance-level
details provided in \supp{4}. Agreement with
Croft et al.\ is 43 of 63 ($\kappa=0.36$). Of its 20 disagreements, 11
are corrections on functions Croft et al.\ consider correctly labelled,
and ten of those carry a recorded commit-level cause such as a mass
refactor, a relabelled sibling function, a whitespace-only change, or
test code; the remaining nine are confirmations, and every one of them is
also a disagreement with Risse et al. One function, BigVul 180692, is rejected by the agent and
called vulnerable by both studies; the rejection rests on a
LeakSanitizer run with a positive control and is kept as a documented
disagreement (\supp{4}). Risse et al.\ judged none of the 39 BigVul and
65 DiverseVul functions they call vulnerable decidable from the function
body alone, so this agreement is reached on the population that reading
in isolation cannot settle.

\subsection{RQ2: Manual review and failure analysis}
\label{sec:review-results}

Table~\ref{tab:review} compares the initial agent decisions with the
final consensus assessments from blinded review of 581 randomly sampled
instances (Section~\ref{sec:review-design}), grouped by the initial
agent outcome. Corrections are supported more often than confirmations
in every dataset, and no group falls below 90\%. PrimeVul's unsupported
confirmations are mostly contradicted (six of nine), whereas all ten
unsupported DiverseVul confirmations have insufficient evidence.

Two cases illustrate attribution and correction errors. In PrimeVul
199823, the fault occurs in a caller before the labelled function
receives the disputed input, so it does not establish that function's
responsibility. DiverseVul instance \path{3313_7cca7e4c0083}, initially assigned
\emph{Label correction}, contains an attributed overflow
on both source sides. The vulnerable-side defect contradicts label
correction; persistence on the fixed side concerns repair effectiveness.

Additional cases expose the source and execution failures addressed by
S1 and S4. BigVul 177956 contains a 23-line extract of a 129-line upstream
function. The omitted region contains the invalid access, and the extract
compiles and runs cleanly. Instances 180307 and 180308 reach the labelled
functions but leave the changed branches unexecuted. In 180472, the labelled
function is \texttt{ipow}, while fault frames belong to
\texttt{print\_object}. These cases distinguish source truncation,
incomplete mechanism execution, and incorrect attribution; further
examples appear in \supp{5}.

\parag{Answer to RQ2.}
Blinded consensus review supports 90.0\% to 92.0\% of sampled
\emph{Confirmed} decisions and 92.6\% to 99.0\% of sampled
\emph{Label correction} decisions. The failure cases trace to
truncated source, unexecuted mechanisms, misattributed faults, or a
defect that persists on both sides, the conditions S1 and S4 exist to
check.

\subsection{RQ3: Outcome distribution}
\label{sec:coverage}

Table~\ref{tab:outcomes} summarizes the post-review outcomes for all
35,849 instances originally labelled vulnerable.

\begin{table}[t]
\centering
\captionsetup{font=footnotesize,skip=4pt}
\caption{Post-review outcome distribution.
The first four numeric columns give counts and percentages of each
column's originally vulnerable population. \emph{Distinct total} counts
byte-identical bodies within each outcome, preserving comments and
whitespace; the bottom value counts each body once across all outcomes. The \emph{Attacked but undecided} row includes the 36 reviewed
instances withheld from both learning classes after manual review.}
\label{tab:outcomes}

\tablebody
\footnotesize
\setlength{\tabcolsep}{2.5pt}
\renewcommand{\arraystretch}{1.0}

\begin{tabularx}{\linewidth}
{@{}L{0.25\linewidth}*{5}{>{\raggedleft\arraybackslash}X}@{}}
\toprule
Outcome & BigVul & PrimeVul & DiverseVul & Total & Distinct total \\
\midrule
Confirmed
& 4,412 (40.48)
& 5,537 (92.22)
& 10,561 (55.74)
& 20,510 (57.21)
& 15,890 \\

Label correction 
& 4,730 (43.39)
& 87 (1.45)
& 2,002 (10.57)
& 6,819 (19.02)
& 6,774 \\

Attacked but undecided
& 1,480 (13.58)
& 374 (6.23)
& 6,127 (32.34)
& 7,981 (22.26)
& 7,777 \\

Not successfully measured
& 278 (2.55)
& 6 (0.10)
& 255 (1.35)
& 539 (1.51)
& 539 \\
\midrule
Overall total
& 10,900 (100.00)
& 6,004 (100.00)
& 18,945 (100.00)
& 35,849 (100.00)
& 30,980 \\
\bottomrule
\end{tabularx}
\end{table}

PrimeVul has the highest confirmation rate (92.22\%), BigVul the largest
label-correction share (43.39\%), and DiverseVul the largest attacked but
undecided share (32.34\%).

\parag{Byte-exact distinct counts.}
Byte-exact deduplication reduces the 35,849 instances to 30,980 distinct
bodies (Table~\ref{tab:outcomes}, last column). Repetition is
concentrated in confirmations: 4,620 of the 4,869 repeated instances are
confirmed, against 45 label corrections. The 15,890 distinct confirmed
bodies are the vulnerable class of the final release.

\parag{Composition of the released benchmark.}
After manual-review filtering and deduplication, the final release
contains 537,591 instances: 15,890 labelled vulnerable (2.96\%) and
521,701 labelled non-vulnerable (97.04\%). Training,
validation, and test splits contain 430,023, 53,772, and 53,796 instances,
approximately 80/10/10. An always-non-vulnerable classifier would achieve
97.04\% accuracy, illustrating the need for complementary detection
metrics.

\parag{Project coverage and concentration.}
The release contains 922 project names after trimming whitespace, or
919 when case is ignored. These are source metadata names, not resolved
repository identities; \texttt{linux} and \texttt{linux-2.6} remain
separate. Table~\ref{tab:release-distributions} lists the largest groups.
\texttt{linux} and \texttt{Chrome} contribute 197,780 instances (36.79\%),
and the ten largest groups contribute 270,026 (50.23\%). This concentration
limits interpreting aggregate scores as equal evidence across projects.

\begin{table}[t]
\caption{Ten largest source project groups and CWE categories in the
release, ranked independently. Project and full-dataset
CWE percentages use 537,591 instances; vulnerable-subset CWE percentages
use 15,890. An instance may have multiple CWE annotations.}
\label{tab:release-distributions}
\tablebody
\small
\setlength{\tabcolsep}{3pt}
\centering
\begin{tabular}{@{}lrr@{\hspace{9pt}}lrr@{\hspace{9pt}}lrr@{}}
\toprule
\multicolumn{3}{c}{Projects: all instances} &
\multicolumn{3}{c}{CWE: all instances} &
\multicolumn{3}{c}{CWE: vulnerable instances} \\
\cmidrule(lr){1-3}\cmidrule(lr){4-6}\cmidrule(l){7-9}
Project & Count & \% & CWE & Count & \% & CWE & Count & \% \\
\midrule
linux       & 123,137 & 22.91 & 119 & 49,574 & 9.22 & 119 & 1,973 & 12.42 \\
Chrome      & 74,643  & 13.88 & 787 & 43,230 & 8.04 & 125 & 1,674 & 10.53 \\
qemu        & 10,760  & 2.00  & 20  & 43,098 & 8.02 & 787 & 1,673 & 10.53 \\
php-src     & 10,602  & 1.97  & 125 & 34,985 & 6.51 & 20  & 1,272 & 8.01 \\
linux-2.6   & 10,197  & 1.90  & 416 & 31,061 & 5.78 & 200 & 830   & 5.22 \\
Android     & 8,885   & 1.65  & 703 & 25,559 & 4.75 & 476 & 813   & 5.12 \\
gpac        & 8,241   & 1.53  & 476 & 23,774 & 4.42 & 703 & 779   & 4.90 \\
server      & 8,092   & 1.51  & 399 & 22,672 & 4.22 & 416 & 700   & 4.41 \\
openssl     & 8,080   & 1.50  & 200 & 21,702 & 4.04 & 399 & 595   & 3.74 \\
ImageMagick & 7,389   & 1.37  & 264 & 21,563 & 4.01 & 190 & 587   & 3.69 \\
\bottomrule
\end{tabular}
\end{table}

\parag{CWE coverage and interpretation.}
Across both classes of the release, 446,823 instances
carry valid CWE annotations spanning 153 categories; 90,768 (16.88\%) lack them. Multiple CWEs occur in 51,249
instances, yielding 503,611 instance--CWE associations, each identifier
counted once per instance. Among the 15,890 retained vulnerable instances,
14,181 have annotations spanning 145 categories and 1,709 (10.76\%) lack them;
2,151 carry multiple CWEs, yielding 16,587 associations.
The three largest categories among vulnerable instances, CWE-119,
CWE-125, and CWE-787 (Table~\ref{tab:release-distributions}), are all
memory-safety weaknesses and together cover a third of the vulnerable
subset.

CWE annotations are source metadata: a non-vulnerable instance carries
the CWE of its source commit, and the defect observed in a confirmed
instance can differ from its CWE (Section~\ref{sec:interpretation}).

\parag{Answer to RQ3.}
The audit yields 20,510 confirmations, 6,819 label corrections, 7,981
attacked-but-undecided instances, and 539 that were not successfully
measured. Confirmation dominates PrimeVul and DiverseVul; label
correction dominates BigVul, where 43.39\% of the instances originally
labelled vulnerable are corrected. The release holds 15,890 distinct
confirmed bodies against 521,701 non-vulnerable instances, and ten
projects supply half of all instances.

\begin{table}[t]
\caption{Vulnerability detection performance on BigVul and DiverseVul.
Each model has five settings, $O/O$, $C/C$, $O/C$, $O/E$, and $C/E$,
where $E$ removes only label-correction instances from the corrected
test set and leaves the remaining instances and labels unchanged. Graph
availability restricts model-specific test populations. Label
combinations and threshold selection follow
Section~\ref{sec:model-design}. Leading zeros are omitted.}
\label{tab:detectors}
\tablebody\footnotesize\setlength{\tabcolsep}{3pt}
\centering
\begin{tabular}{@{}lcc rrrrrr rrrrrr@{}}
\toprule
\multirow{2}{*}{Model} & \multirow{2}{*}{Train} & \multirow{2}{*}{Test} &
\multicolumn{6}{c}{BigVul} & \multicolumn{6}{c@{}}{DiverseVul} \\
\cmidrule(lr){4-9} \cmidrule(l){10-15}
 & & & ACC & Prec. & Rec. & F1 & MCC & AUROC
     & ACC & Prec. & Rec. & F1 & MCC & AUROC \\
\midrule
\multirow{5}{*}{CodeT5}
 & O & O & .9878 & .9440 & .8303 & .8835 & .8791 & .9827 & .8962 & .2570 & .4369 & .3236 & .2827 & .8257 \\
 & C & C & .9832 & .7086 & .7007 & .7047 & .6961 & .9800 & .9170 & .1607 & .3658 & .2233 & .2042 & .8162 \\
 & O & C & .9697 & .4817 & .8309 & .6098 & .6193 & .9719 & .9064 & .1623 & .4490 & .2385 & .2299 & .8301 \\
 & O & E & .9891 & .8040 & .8309 & .8172 & .8117 & .9840 & .9083 & .1667 & .4490 & .2432 & .2340 & .8319 \\
 & C & E & .9886 & .8850 & .7007 & .7822 & .7820 & .9846 & .9172 & .1623 & .3658 & .2248 & .2054 & .8166 \\
\midrule
\multirow{5}{*}{LineVul}
 & O & O & .9847 & .9074 & .8085 & .8551 & .8487 & .9646 & .9008 & .2591 & .4007 & .3147 & .2712 & .8184 \\
 & C & C & .9809 & .6660 & .6636 & .6648 & .6550 & .9504 & .9265 & .1854 & .3686 & .2467 & .2264 & .8215 \\
 & O & C & .9682 & .4670 & .8160 & .5940 & .6033 & .9599 & .9114 & .1580 & .3960 & .2259 & .2106 & .8217 \\
 & O & E & .9870 & .7569 & .8160 & .7853 & .7792 & .9713 & .9134 & .1629 & .3960 & .2309 & .2150 & .8238 \\
 & C & E & .9845 & .7744 & .6636 & .7147 & .7090 & .9544 & .9267 & .1871 & .3686 & .2482 & .2276 & .8219 \\
\midrule
\multirow{5}{*}{DeepDFA}
 & O & O & .9732 & .8120 & .5817 & .6778 & .6743 & .8408 & .7186 & .1158 & .5730 & .1927 & .1557 & .7256 \\
 & C & C & .9782 & .5747 & .7386 & .6464 & .6407 & .9122 & .9276 & .1758 & .3001 & .2217 & .1938 & .8018 \\
 & O & C & .9785 & .5856 & .6911 & .6340 & .6252 & .8987 & .7227 & .0736 & .6101 & .1314 & .1360 & .7428 \\
 & O & E & .9835 & .7022 & .6911 & .6966 & .6881 & .9021 & .7236 & .0742 & .6101 & .1324 & .1368 & .7436 \\
 & C & E & .9837 & .6907 & .7386 & .7139 & .7059 & .9159 & .9279 & .1779 & .3001 & .2234 & .1953 & .8025 \\
\midrule
\multirow{5}{*}{Devign}
 & O & O & .9204 & .2722 & .2685 & .2703 & .2282 & .6954 & .8765 & .1889 & .3560 & .2468 & .1975 & .7408 \\
 & C & C & .9471 & .2051 & .2782 & .2361 & .2119 & .7326 & .9213 & .1533 & .3119 & .2056 & .1811 & .7752 \\
 & O & C & .9448 & .1949 & .2805 & .2300 & .2059 & .7049 & .8903 & .1226 & .3837 & .1859 & .1708 & .7593 \\
 & O & E & .9483 & .2194 & .2805 & .2462 & .2217 & .7089 & .8906 & .1238 & .3837 & .1872 & .1719 & .7601 \\
 & C & E & .9469 & .2104 & .2782 & .2396 & .2149 & .7328 & .9214 & .1545 & .3119 & .2066 & .1819 & .7755 \\
\midrule
\multirow{5}{*}{ReVeal}
 & O & O & .9223 & .3151 & .3547 & .3337 & .2932 & .7294 & .8731 & .1797 & .3454 & .2364 & .1858 & .7037 \\
 & C & C & .9315 & .1510 & .2874 & .1979 & .1752 & .6486 & .9125 & .1396 & .3251 & .1953 & .1726 & .7332 \\
 & O & C & .9494 & .2030 & .2460 & .2225 & .1975 & .7135 & .8878 & .1202 & .3856 & .1832 & .1685 & .7260 \\
 & O & E & .9537 & .2388 & .2460 & .2424 & .2185 & .7187 & .8879 & .1210 & .3856 & .1842 & .1692 & .7264 \\
 & C & E & .9321 & .1570 & .2874 & .2031 & .1794 & .6492 & .9126 & .1405 & .3251 & .1962 & .1733 & .7333 \\
\bottomrule
\end{tabular}
\end{table}

\subsection{RQ4: Vulnerability detector performance changes after label correction}
\label{sec:model-results}

Table~\ref{tab:detectors} reports five detectors under five settings
on BigVul and DiverseVul. PrimeVul's detector results appear in
the supplementary material.

\parag{Overall changes ($O/O$ to $C/C$).}
F1 decreases for all five detectors on BigVul and four on
DiverseVul. DeepDFA is the exception on DiverseVul, improving
from 0.1927 to 0.2217. These comparisons combine changes in
training data, evaluation labels, and retained populations.

\parag{Re-evaluating model without retraining ($O/O$ to $O/C$).}
F1, precision, and MCC decrease for every detector on both
datasets. The F1 losses range from 0.0403 to 0.2737 on BigVul
and from 0.0532 to 0.0888 on DiverseVul. At LineVul's fixed
threshold of 0.5 on BigVul, precision falls from 0.9074 to
0.4670, while AUROC changes from 0.9646 to 0.9599.
These results show different metric responses to revised
reference labels and evaluation populations. BigVul's graph-model
$O/C$ results additionally use reselected validation thresholds
(Section~\ref{sec:model-design}); DiverseVul retains the original
thresholds.

\parag{Retraining with corrected labels ($O/C$ to $C/C$).}
F1 and MCC improve for four of five detectors on each dataset.
The exceptions are ReVeal on BigVul and CodeT5 on DiverseVul.
CodeT5 has the largest BigVul F1 gain (0.0949), while DeepDFA
has the largest DiverseVul gain (0.0903). Responses vary across
models and datasets, including cases where higher F1 coexists
with lower AUROC.

\parag{Withholding the relabelled instances ($O/E$ versus $C/E$).}
For each trained model, removing correction-derived negatives at
a fixed threshold leaves recall unchanged and makes F1
nondecreasing. This filtering effect alone does not establish
label correctness. The informative comparison is whether the
corrected-training advantage persists on the common test view.
On DiverseVul, corrected training retains higher F1 and MCC
for four detectors. On BigVul, $C/E$ F1 is lower than $O/E$
for CodeT5, LineVul, Devign, and ReVeal, while DeepDFA improves
from 0.6966 to 0.7139. Relative to the comparisons on $C$,
the gain changes sign for CodeT5, LineVul, and Devign;
ReVeal remains worse under both views. Devign's difference
on $E$ is small (0.0066).

\parag{Answer to RQ4.}
Corrected evaluation protocols lower reported F1, while retraining
gains depend on the dataset and the treatment of label-correction
instances. All training comparisons are single-run point estimates;
their stability across training runs remains unmeasured.

\begin{table}[t]
\caption{Test performance on the merged benchmark constructed from
BigVul, PrimeVul, and DiverseVul. Threshold selection follows
Section~\ref{sec:model-design}. Leading zeros are omitted.}
\label{tab:combined}
\tablebody
\centering
\begin{tabular}{@{}lrrrrrr@{}}
\toprule
Model & ACC & Prec. & Rec. & F1 & MCC & AUROC \\
\midrule
CodeT5  & .9625 & .4471 & .3966 & .4203 & .4018 & .9125 \\
LineVul & .9654 & .4888 & .1688 & .2509 & .2734 & .8423 \\
DeepDFA & .9236 & .1563 & .2541 & .1936 & .1609 & .7559 \\
ReVeal  & .9342 & .1840 & .2422 & .2091 & .1773 & .7858 \\
Devign  & .9462 & .2452 & .2399 & .2425 & .2146 & .7371 \\
\bottomrule
\end{tabular}
\end{table}

\subsection{Vulnerability detection performance on the merged dataset}
\label{sec:combined-results}

Table~\ref{tab:combined} reports models trained or fine-tuned on corrected
data merged from BigVul, PrimeVul, and DiverseVul using byte-exact
deduplication and the audit decisions fixed for these runs. The policy
in Section~\ref{sec:dataset-construction} addresses identified label
errors and conflicts while omitting unresolved instances, providing more
consistent training and evaluation labels.

\parag{Transformer-based detectors.}
CodeT5 leads LineVul by 0.17 in F1, 0.13 in MCC, and 0.07 in AUROC. The
F1 gap is a recall gap (0.3966 versus 0.1688) at similar precision
(0.4471 versus 0.4888).

\parag{Graph-based detectors and cross-family comparison.}
Among the graph models, Devign has the best F1 and MCC and ReVeal the
best AUROC; DeepDFA trades the highest recall for the lowest precision
and F1. Both transformer models exceed the graph models on F1, MCC, and
AUROC, but LineVul's F1 lead over Devign is 0.0084 and its recall is the
lowest of the five. Graph preprocessing restricts the evaluated
functions, while text models use the full text input population, so the
cross-family comparison is not on a common function set.

\parag{Class imbalance and implications for RQ4.}
Accuracy above 0.96 for the transformer models reflects the 97.04\%
non-vulnerable majority, not detection: the best detector, CodeT5, finds
39.66\% of the vulnerable functions, and 44.71\% of its positive
predictions are correct. The merged benchmark removes the contradictory
supervision described in the Introduction; on it, CodeT5 leads on F1,
MCC, and AUROC, and every detector leaves most vulnerable functions
undetected.

\section{Threats to Validity}
\label{sec:discussion}

\parag{Construct.}
\emph{Confirmed} asserts an attributed defect in the historical function
under the chosen oracle, not that it is the CVE's defect, that the fix
removes it, or that an attacker can reach it
(Section~\ref{sec:interpretation}); this is where the agent and Risse et
al.\ part ways (Section~\ref{sec:external-results}). \emph{Label correction} maps to evidence-supported empirical
non-vulnerable labels ($y=0$) under all seven checks,
source-grounded rejection rationale $R$, and the retention policy
(Section~\ref{sec:dataset-construction}). Blinded review supported
272/281 sampled initial corrections (96.8\%; Table~\ref{tab:review}).
This pooled sample support rate does not estimate residual vulnerability
prevalence; undetected defects may still cause mislabelling.

\parag{Internal.}
Each audit decision was produced by one run of a single agent
configuration, so decision stability across agent runs or model
configurations was not evaluated. The manual review was conducted by
two of the authors; blinding withheld the agent's decisions during
individual assessment and consensus discussion, while the reviewers
were familiar with the S6 criteria that they helped define. With
approximately 100 reviewed instances per outcome group, the reported
support rates remain subject to sampling uncertainty; for example, a
90\% support rate corresponds to a 95\% confidence interval of roughly
83\%--94\%. The external comparison includes only functions for which
both sides reached a decision, and neither prior manual labelling is
treated as ground truth. Comparisons from $O/O$ to $C/C$ jointly change
labels, sample membership, deduplication, and class proportions, so the
observed performance differences reflect their combined effects.
Each training setting is represented by one run; the reported F1
differences are therefore descriptive, while their stability across
training runs remains unassessed. Residual errors in retained original
non-vulnerable labels may also affect measured precision.

\parag{External and reproducibility.}
The audit covers C and C++ functions from three datasets, and
\texttt{linux} and \texttt{Chrome} supply 36.79\% of the release.
Re-auditing costs about 20 minutes of agent time per instance
(USD~60,000 in total); reproducing a decision does not, since each
experiment replays from its Docker image, with gaps listed in \supp{8}.

\section{Related Work}
\label{sec:related}

\parag{Vulnerability datasets and label quality.}
BigVul and DiverseVul derive function labels from historical vulnerabilities
and fixing commits~\cite{fan2020bigvul,chen2023diversevul}. PrimeVul improves
labelling and evaluation procedures~\cite{ding2025primevul}.
CVEfixes collects vulnerabilities and fixes, while D2A derives labels
through differential static analysis~\cite{bhandari2021cvefixes,zheng2021d2a}.
Croft et al.\ examine label accuracy and duplication, and Risse et al.\
identify incorrect fixing commits and unrelated
changes~\cite{croft2023dataquality,risse2025wrongexam}.
CleanVul uses LLM-based filtering of vulnerability-fixing
commits~\cite{li2024cleanvul}. \vulvalidate complements these approaches
with dynamic evidence for function-level label correction and conflict
resolution, while retaining unresolved instances.

\parag{Executable vulnerability reproduction and evidence.}
ARVO reconstructs historical OSS-Fuzz vulnerabilities, replays triggering
inputs on vulnerable and fixed revisions, locates repairs, and provides
reusable environments~\cite{mei2026arvo}. CVE-Factory constructs executable
vulnerability tasks and evaluates source authenticity, dynamic execution,
and repair validity~\cite{luo2026cvefactory}. Magma distinguishes reaching,
triggering, and detecting defects~\cite{hazimeh2020magma}, while fuzzing
studies explain how unsuccessful searches depend on inputs, budgets, and
randomness~\cite{klees2018evaluating}. \vulvalidate binds the tested function
to historical source, checks execution of the relevant operation, and
establishes function-level attribution. Label correction requires
mechanism-relevant evidence and a source-grounded rejection rationale,
with attack histories preserving the tested scope.

\parag{LLM agents and feedback-guided testing.}
SWE-agent and AutoCodeRover combine code navigation, modification, and
tool execution~\cite{yang2024sweagent,zhang2024autocoderover}.
ChatAFL uses LLMs for protocol fuzzing~\cite{meng2024chatafl}, and
VulTrigger combines program analysis with agent-generated downstream
tests for third-party library vulnerabilities~\cite{ju7275148vultrigger}.
AFL++ and directed fuzzing guide exploration using execution
feedback~\cite{fioraldi2020aflpp,bohme2017aflgo}. \vulvalidate uses agents
to orchestrate dynamic tools, construct historical experiments, and refine
inputs and harnesses, with decisions governed by the S6 contract and
checked by blinded two-expert review.

\parag{Learning-based vulnerability detection and evaluation.}
Devign and ReVeal learn graph-based representations, while DeepDFA uses
abstract dataflow information~\cite{zhou2019devign,chakraborty2022reveal,steenhoek2024deepdfa}.
LineVul and CodeT5 use pretrained transformers for vulnerability or defect
detection~\cite{fu2022linevul,wang2021codet5}.
Empirical studies examine how data quality and experimental choices affect
detector performance~\cite{jimenez2019labelling,nong2022openscience,steenhoek2023empirical}.
Our experiments distinguish re-evaluation without retraining from retraining
with corrected labels. The corrected merged benchmark extends evaluation to combined
data after resolving identified label conflicts.

\section{Conclusion}
\label{sec:conclusion}

\vulvalidate uses LLM agents to audit function-level vulnerability labels
through reproducible dynamic experiments. It combines historical source
reconstruction, feedback-guided trigger construction, mechanism-specific
oracles, and controlled execution of vulnerable and fixed revisions
with function-level attribution. The audit covers 35,849 instances
originally labelled vulnerable in BigVul, PrimeVul, and DiverseVul,
yielding 20,510 confirmations and 6,819 label corrections and leaving
8,520 instances unresolved. The release contains 15,890 distinct
confirmed vulnerable function bodies after byte-exact deduplication. The merged benchmark
resolves identified label conflicts under the construction policy and
provides more consistent training and evaluation data.

Blinded review of 581 randomly sampled instances by two experts supports
90.0\% to 92.0\% of confirmations and 92.6\% to 99.0\% of label
corrections, and agreement with independently produced manual labels yields
Cohen's $\kappa$ values from 0.36 to 0.55, providing an external
reference for the audit decisions.
Re-evaluating models trained on the original labels against corrected labels lowers F1 for
every detector, and retraining on corrected labels recovers part of the
loss for four of five detectors on each dataset. The merged
benchmark further shows that high accuracy or AUROC can coexist with
limited recall on imbalanced data. These results connect label quality
to the interpretation of detector performance. The corrected benchmark,
executable evidence, and reusable \vulvalidate skill provide a basis for
more reliable vulnerability detection research, with label decisions
tied to their supporting evidence and tested conditions.

\bibliographystyle{ACM-Reference-Format}
\bibliography{references}

@inproceedings{fan2020bigvul,
  title={A C/C++ code vulnerability dataset with code changes and CVE summaries},
  author={Fan, Jiahao and Li, Yi and Wang, Shaohua and Nguyen, Tien N},
  booktitle={Proceedings of the 17th international conference on mining software repositories},
  pages={508--512},
  year={2020}
}

@inproceedings{zhou2019devign,
 title={Devign: Effective vulnerability identification by learning comprehensive program semantics via graph neural networks},
  author={Zhou, Yaqin and Liu, Shangqing and Siow, Jingkai and Du, Xiaoning and Liu, Yang},
  booktitle={Advances in neural information processing systems},
  volume={32},
  year={2019}
}

@inproceedings{chen2023diversevul,
  title={DiverseVul: A New Vulnerable Source Code Dataset for Deep Learning Based Vulnerability Detection},
  author={Chen, Yizheng and Ding, Zhoujie and Alowain, Lamya and Chen, Xinyun and Wagner, David},
  booktitle={Proceedings of the 26th international symposium on research in attacks, intrusions and defenses},
  pages={654--668},
  year={2023}
}

@misc{ding2025primevul,
  title={Vulnerability detection with code language models: How far are we?},
  author={Ding, Yangruibo and Fu, Yanjun and Ibrahim, Omniyyah and Sitawarin, Chawin and Chen, Xinyun and Alomair, Basel and Wagner, David and Ray, Baishakhi and Chen, Yizheng},
  howpublished={arXiv preprint arXiv:2403.18624},
  year={2024}
}

@inproceedings{croft2023dataquality,
  title={Data quality for software vulnerability datasets},
  author={Croft, Roland and Babar, M Ali and Kholoosi, M Mehdi},
  booktitle={2023 IEEE/ACM 45th International Conference on Software Engineering (ICSE)},
  pages={121--133},
  year={2023},
  organization={IEEE}
}

@article{risse2025wrongexam,
  title={Top score on the wrong exam: On benchmarking in machine learning for vulnerability detection},
  author={Risse, Niklas and Liu, Jing and B{\"o}hme, Marcel},
  journal={Proceedings of the ACM on Software Engineering},
  volume={2},
  number={ISSTA},
  pages={388--410},
  year={2025},
  publisher={ACM New York, NY, USA}
}

@article{chakraborty2022reveal,
  title={Deep learning based vulnerability detection: Are we there yet?},
  author={Chakraborty, Saikat and Krishna, Rahul and Ding, Yangruibo and Ray, Baishakhi},
  journal={IEEE Transactions on Software Engineering},
  volume={48},
  number={9},
  pages={3280--3296},
  year={2021},
  publisher={IEEE}
}

@inproceedings{fu2022linevul,
  title={LineVul: A Transformer-based Line-Level Vulnerability Prediction},
  author={Fu, Michael and Tantithamthavorn, Chakkrit},
  booktitle={Proceedings of the 19th international conference on mining software repositories},
  pages={608--620},
  year={2022}
}

@inproceedings{steenhoek2024deepdfa,
  title={Dataflow analysis-inspired deep learning for efficient vulnerability detection},
  author={Steenhoek, Benjamin and Gao, Hongyang and Le, Wei},
  booktitle={Proceedings of the 46th IEEE/ACM International Conference on Software Engineering},
  pages={1--13},
  year={2024}
}

@inproceedings{serebryany2012asan,
 title={AddressSanitizer: A fast address sanity checker},
  author={Serebryany, Konstantin and Bruening, Derek and Potapenko, Alexander and Vyukov, Dmitriy},
  booktitle={2012 USENIX annual technical conference (USENIX ATC 12)},
  pages={309--318},
  year={2012}
}

@inproceedings{stepanov2015msan,
 title={MemorySanitizer: fast detector of uninitialized memory use in C++},
  author={Stepanov, Evgeniy and Serebryany, Konstantin},
  booktitle={2015 IEEE/ACM International Symposium on Code Generation and Optimization (CGO)},
  pages={46--55},
  year={2015},
  organization={IEEE}
}

@inproceedings{serebryany2009tsan,
  title={ThreadSanitizer: data race detection in practice},
  author={Serebryany, Konstantin and Iskhodzhanov, Timur},
  booktitle={Proceedings of the workshop on binary instrumentation and applications},
  pages={62--71},
  year={2009}
}

@article{nethercote2007valgrind,
 title={Valgrind: a framework for heavyweight dynamic binary instrumentation},
  author={Nethercote, Nicholas and Seward, Julian},
  journal={ACM Sigplan notices},
  volume={42},
  number={6},
  pages={89--100},
  year={2007},
  publisher={ACM New York, NY, USA}
}

@misc{llvm2026ubsan,
  author = {{LLVM Project}},
  title = {UndefinedBehaviorSanitizer},
  year = {2026},
  howpublished = {Clang documentation},
  url = {https://clang.llvm.org/docs/UndefinedBehaviorSanitizer.html},
  note = {Accessed 2026-09-17}
}

@misc{llvm2026lsan,
  author = {{LLVM Project}},
  title = {LeakSanitizer},
  year = {2026},
  howpublished = {Clang documentation},
  url = {https://clang.llvm.org/docs/LeakSanitizer.html},
  note = {Accessed 2026-09-17}
}

@misc{linux2026kasan,
  author = {{Linux Kernel Documentation}},
  title = {Kernel Address Sanitizer},
  year = {2026},
  url = {https://docs.kernel.org/dev-tools/kasan.html},
  note = {Accessed 2026-09-17}
}

@misc{linux2026kmsan,
  author = {{Linux Kernel Documentation}},
  title = {Kernel Memory Sanitizer},
  year = {2026},
  url = {https://docs.kernel.org/dev-tools/kmsan.html},
  note = {Accessed 2026-09-17}
}

@misc{linux2026kcsan,
  author = {{Linux Kernel Documentation}},
  title = {Kernel Concurrency Sanitizer},
  year = {2026},
  url = {https://docs.kernel.org/dev-tools/kcsan.html},
  note = {Accessed 2026-09-17}
}

@inproceedings{fioraldi2020aflpp,
  title={{AFL++}: Combining incremental steps of fuzzing research},
  author={Fioraldi, Andrea and Maier, Dominik and Ei{\ss}feldt, Heiko and Heuse, Marc},
  booktitle={14th USENIX workshop on offensive technologies (WOOT 20)},
  year={2020}
}

@inproceedings{bellard2005qemu,
  title={QEMU, a fast and portable dynamic translator},
  author={Bellard, Fabrice},
  booktitle={2005 USENIX Annual Technical Conference (USENIX ATC 05)},
  pages={41--46},
  year={2005},
  publisher={USENIX Association},
  url={https://www.usenix.org/conference/2005-usenix-annual-technical-conference/qemu-fast-and-portable-dynamic-translator}
}

@inproceedings{kivity2007kvm,
  title={kvm: the Linux virtual machine monitor},
  author={Kivity, Avi and Kamay, Yaniv and Laor, Dor and Lublin, Uri and Liguori, Anthony},
  booktitle={Proceedings of the Linux symposium},
  volume={1},
  pages={225--230},
  year={2007},
  address={Ottawa, Ontario, Canada},
  url={https://www.kernel.org/doc/ols/2007/ols2007v1-pages-225-230.pdf}
}

@article{merkel2014docker,
  title={Docker: lightweight linux containers for consistent development and deployment},
  author={Merkel, Dirk},
  journal={Linux Journal},
  volume={239},
  number={2},
  pages={2},
  year={2014},
  url={https://www.linuxjournal.com/content/docker-lightweight-linux-containers-consistent-development-and-deployment}
}

@misc{anthropic2026claudecode,
  author = {{Anthropic}},
  title = {Claude Code Documentation},
  year = {2026},
  url = {https://code.claude.com/docs/en/overview},
  note = {Accessed 2026-09-17}
}

@misc{anthropic2026opus5,
  author = {{Anthropic}},
  title = {Introducing {Claude Opus 5}},
  year = {2026},
  month = jul,
  url = {https://www.anthropic.com/news/claude-opus-5},
  note = {Accessed 2026-09-17}
}

@article{hazimeh2020magma,
 title={Magma: A ground-truth fuzzing benchmark},
  author={Hazimeh, Ahmad and Herrera, Adrian and Payer, Mathias},
  journal={Proceedings of the ACM on Measurement and Analysis of Computing Systems},
  volume={4},
  number={3},
  pages={1--29},
  year={2020},
  publisher={ACM New York, NY, USA}
}

@inproceedings{bohme2017aflgo,
  title={Directed greybox fuzzing},
  author={B{\"o}hme, Marcel and Pham, Van-Thuan and Nguyen, Manh-Dung and Roychoudhury, Abhik},
  booktitle={Proceedings of the 2017 ACM SIGSAC conference on computer and communications security},
  pages={2329--2344},
  year={2017}
}

@inproceedings{klees2018evaluating,
  title={Evaluating fuzz testing},
  author={Klees, George and Ruef, Andrew and Cooper, Benji and Wei, Shiyi and Hicks, Michael},
  booktitle={Proceedings of the 2018 ACM SIGSAC conference on computer and communications security},
  pages={2123--2138},
  year={2018}
}

@inproceedings{osterlund2020parmesan,
  title = {{ParmeSan}: Sanitizer-Guided Greybox Fuzzing},
  author={{\"O}sterlund, Sebastian and Razavi, Kaveh and Bos, Herbert and Giuffrida, Cristiano},
  booktitle={29th USENIX Security Symposium (USENIX Security 20)},
  pages={2289--2306},
  year={2020}
}

@inproceedings{meng2024chatafl,
  title={Large Language Model guided Protocol Fuzzing},
  author={Meng, Ruijie and Mirchev, Martin and B{\"o}hme, Marcel and Roychoudhury, Abhik},
  booktitle={NDSS},
  year={2024}
}

@inproceedings{yang2024sweagent,
  title={SWE-agent: Agent-Computer Interfaces Enable Automated Software Engineering},
 author={Yang, John and Jimenez, Carlos and Wettig, Alexander and Lieret, Kilian and Yao, Shunyu and Narasimhan, Karthik and Press, Ofir},
  journal={Advances in Neural Information Processing Systems},
  volume={37},
  pages={50528--50652},
  year={2024}
}

@inproceedings{zhang2024autocoderover,
  title={AutoCodeRover: Autonomous Program Improvement},
  author={Zhang, Yuntong and Ruan, Haifeng and Fan, Zhiyu and Roychoudhury, Abhik},
  booktitle={Proceedings of the 33rd ACM SIGSOFT International Symposium on Software Testing and Analysis},
  pages={1592--1604},
  year={2024}
}

@inproceedings{jimenez2024swebench,
  title={SWE-bench: Can Language Models Resolve Real-world Github Issues?},
  author={Jimenez, Carlos E and Yang, John and Wettig, Alexander and Yao, Shunyu and Pei, Kexin and Press, Ofir and Narasimhan, Karthik},
  booktitle={International Conference on Learning Representations},
  year={2024}
}

@misc{luo2026cvefactory,
  title={CVE-Factory: Scaling Expert-Level Agentic Tasks for Code Security Vulnerability},
  author={Luo, Xianzhen and Zhang, Jingyuan and Zhou, Shiqi and Huang, Jinyang and Xiao, Chuan and Zhu, Qingfu and Ma, Zhiyuan and Yue, Xing and Yue, Yang and Zeng, Wencong and others},
  howpublished={arXiv preprint arXiv:2602.03012},
  year={2026}
}

@inproceedings{bhandari2021cvefixes,
 title={CVEfixes: automated collection of vulnerabilities and their fixes from open-source software},
  author={Bhandari, Guru and Naseer, Amara and Moonen, Leon},
  booktitle={Proceedings of the 17th international conference on predictive models and data analytics in software engineering},
  pages={30--39},
  year={2021}
}

@inproceedings{zheng2021d2a,
  title={D2A: A Dataset Built for AI-Based Vulnerability Detection Methods Using Differential Analysis},
  author={Zheng, Yunhui and Pujar, Saurabh and Lewis, Burn and Buratti, Luca and Epstein, Edward and Yang, Bo and Laredo, Jim and Morari, Alessandro and Su, Zhong},
  booktitle={2021 IEEE/ACM 43rd International Conference on Software Engineering: Software Engineering in Practice (ICSE-SEIP)},
  pages={111--120},
  year={2021},
  organization={IEEE}
}

@article{barr2015oracle,
  title={The oracle problem in software testing: A survey},
  author={Barr, Earl T and Harman, Mark and McMinn, Phil and Shahbaz, Muzammil and Yoo, Shin},
  journal={IEEE transactions on software engineering},
  volume={41},
  number={5},
  pages={507--525},
  year={2015},
  publisher={IEEE},
  doi={10.1109/TSE.2014.2372785}
}

@inproceedings{mei2026arvo,
  title={ARVO: Atlas of reproducible vulnerabilities for open-source software},
  author={Mei, Xiang and Del Castillo, Jordi and Singaria, Pulkit Singh and Xi, Haoran and Benchikh, Abdelouahab and Bao, Tiffany and Wang, Ruoyu and Shoshitaishvili, Yan and Doup{\'e}, Adam and Pearce, Hammond and others},
  booktitle={2026 IEEE 11th European Symposium on Security and Privacy (EuroS\&P)},
  pages={860--873},
  year={2026},
  organization={IEEE}
}

@article{ju7275148vultrigger,
  title={VulTrigger: Multi-Agent LLM Generation of Downstream Vulnerability-Triggering Tests for Third-Party Libraries},
  author={JU, Xiaolin and Yu, Xinhua and Wang, Rongcun and Li, Chang and Chen, Xiang},
  journal={Available at SSRN 7275148},
  year={2026}
}

@misc{google2026ossfuzz,
  author = {{Google}},
  title = {{OSS-Fuzz}: Continuous Fuzzing for Open Source Software},
  year = {2026},
  url = {https://google.github.io/oss-fuzz/},
  note = {Accessed 2026-09-25}
}

@misc{googleossfuzztargets,
  author = {{Google Open Source Security Team}},
  title = {Fuzz Target Generation Using {LLMs}},
  howpublished = {OSS-Fuzz research notes},
  year = {2023},
  url = {https://google.github.io/oss-fuzz/research/llms/target_generation/},
  note = {Accessed 2026-09-25}
}

@inproceedings{zhang2026coderover,
   title={Fixing Security Vulnerabilities with Agentic AI in OSS-Fuzz},
  author={Zhang, Yuntong and Wang, Jiawei and Berzin, Dominic and Mirchev, Martin and Roychoudhury, Abhik},
  booktitle={Proceedings of the IEEE/ACM 48th International Conference on Software Engineering: Software Engineering in Practice},
  pages={374--384},
  year={2026}
}

@inproceedings{jimenez2019labelling,
  title={The importance of accounting for real-world labelling when predicting software vulnerabilities},
  author={Jimenez, Matthieu and Rwemalika, Renaud and Papadakis, Mike and Sarro, Federica and Le Traon, Yves and Harman, Mark},
  booktitle={Proceedings of the 2019 27th ACM Joint Meeting on European Software Engineering Conference and Symposium on the Foundations of Software Engineering},
  pages={695--705},
  year={2019}
}

@article{nong2022openscience,
  title={Open science in software engineering: A study on deep learning-based vulnerability detection},
  author={Nong, Yu and Sharma, Rainy and Hamou-Lhadj, Abdelwahab and Luo, Xiapu and Cai, Haipeng},
  journal={IEEE Transactions on Software Engineering},
  volume={49},
  number={4},
  pages={1983--2005},
  year={2022},
  publisher={IEEE}
}

@inproceedings{steenhoek2023empirical,
  title={An empirical study of deep learning models for vulnerability detection},
  author={Steenhoek, Benjamin and Rahman, Md Mahbubur and Jiles, Richard and Le, Wei},
  booktitle={2023 IEEE/ACM 45th International Conference on Software Engineering (ICSE)},
  pages={2237--2248},
  year={2023},
  organization={IEEE}
}

@misc{li2024cleanvul,
  title={Cleanvul: Automatic function-level vulnerability detection in code commits using llm heuristics},
  author={Li, Yikun and Zhang, Ting and Widyasari, Ratnadira and Tun, Yan Naing and Nguyen, Huu Hung and Bui, Tan and Irsan, Ivana Clairine and Cheng, Yiran and Lan, Xiang and Ang, Han Wei and others},
  journal={arXiv preprint arXiv:2411.17274},
  year={2024}
}

@inproceedings{wang2021codet5,
  title={CodeT5: Identifier-aware unified pre-trained encoder-decoder models for code understanding and generation},
  author={Wang, Yue and Wang, Weishi and Joty, Shafiq and Hoi, Steven CH},
  booktitle={Proceedings of the 2021 conference on empirical methods in natural language processing},
  pages={8696--8708},
  year={2021}
}

@inproceedings{feng2020codebert,
  title={CodeBERT: A pre-trained model for programming and natural languages},
  author={Feng, Zhangyin and Guo, Daya and Tang, Duyu and Duan, Nan and Feng, Xiaocheng and Gong, Ming and Shou, Linjun and Qin, Bing and Liu, Ting and Jiang, Daxin and others},
  booktitle={Findings of the association for computational linguistics: EMNLP 2020},
  pages={1536--1547},
  year={2020}
}

@article{matthews1975comparison,
  title={Comparison of the predicted and observed secondary structure of T4 phage lysozyme},
  author={Matthews, Brian W},
  journal={Biochimica et Biophysica Acta (BBA)-Protein Structure},
  volume={405},
  number={2},
  pages={442--451},
  year={1975},
  publisher={Elsevier}
}

@article{fawcett2006introduction,
  title={An introduction to ROC analysis},
  author={Fawcett, Tom},
  journal={Pattern recognition letters},
  volume={27},
  number={8},
  pages={861--874},
  year={2006},
  publisher={Elsevier}
}

@article{chicco2020advantages,
  title={The advantages of the Matthews correlation coefficient (MCC) over F1 score and accuracy in binary classification evaluation},
  author={Chicco, Davide and Jurman, Giuseppe},
  journal={BMC genomics},
  volume={21},
  number={1},
  pages={6},
  year={2020},
  publisher={Springer}
}

@article{saito2015precision,
  title={The precision-recall plot is more informative than the ROC plot when evaluating binary classifiers on imbalanced datasets},
  author={Saito, Takaya and Rehmsmeier, Marc},
  journal={PloS one},
  volume={10},
  number={3},
  pages={e0118432},
  year={2015},
  publisher={Public Library of Science San Francisco, CA USA}
}

@inproceedings{yao2020assessing,
  title={Assessing software defection prediction performance: Why using the Matthews correlation coefficient matters},
  author={Yao, Jingxiu and Shepperd, Martin},
  booktitle={Proceedings of the 24th international conference on evaluation and assessment in software engineering},
  pages={120--129},
  year={2020}
}

@misc{google2024aifuzzing,
  author       = {{Google}},
  title        = {Leveling Up Fuzzing: Finding More Vulnerabilities with {AI}},
  year         = {2024},
  month        = nov,
  howpublished = {Google Security Blog},
  url          = {https://security.googleblog.com/2024/11/leveling-up-fuzzing-finding-more.html},
  note         = {Accessed 2026-10-01}
}

@misc{darpa2025aixccofficial,
  author       = {{Defense Advanced Research Projects Agency}},
  title        = {{AI} Cyber Challenge Marks Pivotal Inflection Point for Cyber Defense},
  year         = {2025},
  month        = aug,
  howpublished = {DARPA},
  url          = {https://www.darpa.mil/news/2025/aixcc-results},
  note         = {Published August 8, 2025. Accessed 2026-10-01 }
}

@misc{nist2026nvdofficial,
  author       = {{National Institute of Standards and Technology}},
  title        = {{NIST} Updates {NVD} Operations to Address Record {CVE} Growth},
  year         = {2026},
  month        = apr,
  howpublished = {NIST},
  url          = {https://www.nist.gov/news-events/news/2026/04/nist-updates-nvd-operations-address-record-cve-growth},
  note         = {Published April 15, 2026; updated April 17, 2026.}
}

@article{cohen1960kappa,
  title={A coefficient of agreement for nominal scales},
  author={Cohen, Jacob},
  journal={Educational and Psychological Measurement},
  volume={20},
  number={1},
  pages={37--46},
  year={1960},
  publisher={Sage}
}

@article{landis1977kappa,
  title={The measurement of observer agreement for categorical data},
  author={Landis, J. Richard and Koch, Gary G.},
  journal={Biometrics},
  volume={33},
  number={1},
  pages={159--174},
  year={1977}
}
\end{document}